\documentclass[12pt]{article}
\usepackage{amsmath}
\usepackage{graphicx}
\usepackage{enumerate}
\usepackage{natbib}
\usepackage{url} 

\usepackage{natbib,graphicx,setspace,lscape,longtable,epsfig}
\usepackage{tabularx}
\usepackage{amsmath,amsthm,amssymb,color}
\usepackage{longtable}
\usepackage{array}
\usepackage{booktabs}
\usepackage{hyperref}
\usepackage{multirow}
\usepackage{epstopdf}
\usepackage{subfigure}
\usepackage{arydshln}
\UseRawInputEncoding
\newcommand{\blind}{0}

\numberwithin{equation}{section}
\renewcommand{\baselinestretch}{1.75}

\def\defby{\stackrel{\mbox{\textrm{\tiny def}}}{=}}

\newtheorem{defi}{\bf Definition}[section]
\newtheorem{theo}{\bf Theorem}[section]

\newtheorem{coll}{\bf Corollary}[section]
\newtheorem{rema}{\bf Remark}[section]
\newtheorem{cond}{\bf Condition}

\makeatletter

\newcommand{\Rmnum}[1]{\expandafter\@slowromancap\romannumeral #1@}
\makeatother

\DeclareMathOperator*{\argmax}{arg\,max}

\begin{document}

\def\spacingset#1{\renewcommand{\baselinestretch}%
{#1}\small\normalsize} \spacingset{1}


\if0\blind
{
  \title{\bf Multifold Cross-Validation Model Averaging for Generalized Additive Partial Linear Models}
  \author{Ze Chen, Jun Liao, Wangli Xu$^{*}$ and Yuhong Yang\thanks{Yuhong Yang (Email: yangx374@umn.edu) is co-corresponding author and Professor, School of Statistics, University of Minnesota, Minneapolis, MN 55455, United States. Wangli Xu (Email: wlxu@ruc.edu.cn) is co-corresponding author and Professor, Center for Applied Statistics and School of Statistics, Renmin University of China, Beijing 100872, China.
  Ze Chen (Email: chze96@163.com) is PH.D student, Center for Applied Statistics and School of Statistics, Renmin University of China, Beijing 100875, China. Jun Liao (Email: junliao@ruc.edu.cn) is Assistant Professor, Center for Applied Statistics and School of Statistics, Renmin University of China, Beijing 100875, China. }}
  \maketitle
} \fi

\if1\blind
{
  \bigskip
  \bigskip
  \bigskip
  \begin{center}
    {\LARGE\bf Multifold Cross-Validation Model Averaging for Generalized Additive Partial Linear Models}
\end{center}
  \medskip
} \fi

\bigskip
\begin{abstract}
Generalized additive partial linear models (GAPLMs) are appealing for model interpretation and prediction. However, for GAPLMs, the  covariates  and the degree of smoothing in the nonparametric parts are often difficult to determine in practice. To address this model selection uncertainty issue, we develop a computationally feasible model averaging (MA) procedure.  The model weights are  data-driven and selected based on multifold cross-validation (CV) (instead of leave-one-out) for computational saving. When all the candidate models are misspecified, we show that the proposed MA estimator for GAPLMs is asymptotically optimal in the sense of achieving the lowest possible Kullback-Leibler loss. In the other scenario where the candidate model set contains at least one correct model, the weights chosen by the multifold CV are asymptotically concentrated on the correct models. As a by-product, we propose a variable importance measure to quantify the importances of the predictors in GAPLMs based on the MA weights. It is shown to  be able to asymptotically identify the variables in the true model.  Moreover, when the number of candidate models is very large, a model screening method is provided. Numerical experiments show the superiority of the proposed MA method over some existing model averaging and selection methods.
\end{abstract}

\noindent%
\vfill

\newpage
\spacingset{1.5} 
\section{Introduction}
\label{sec:intro}

Generalized additive partial linear models (GAPLMs) offer a powerful tool for studying possibly complicated relationship between a response and some predictors. In GAPLMs, some predictors are modeled as having nonparametric effects, and the remaining predictors affect the response in a linear fashion \citep{Hardle:Huet:Sperlich:2004}. Hence, GAPLMs balance the interpretability of typical parametric models (e.g., generalized linear models (GLMs)) and the flexibility of purely nonparametric models (e.g., generalized additive models). GAPLMs have been widely used in many fields, such as AIDS clinical trials \citep{Shiboski:1998} and credit scoring \citep{Muller:Ronz:2000}.

However, the choices of the covariates  and the degree of smoothing in nonparametric parts of GAPLMs are typically uncertain in practice. Model selection attempts to choose a model with certain covariates and degrees of smoothing. The commonly used model selection criteria include the Akaike information criterion (AIC; \cite{akaike:1973a}), the Bayesian information criterion (BIC; \cite{Schwarz:1978}), and the Mallows $C_p$ criterion \citep{Mallows:1973}, among others. Model selection, however, ignores possibly high uncertainty since it chooses only one model in the selection process (e.g., \cite{Draper:1995,Yuan:Yang:2005}). As an alternative to model selection, model averaging (MA) reduces the instability in model selection and can significantly reduce the statistical risk  \citep{Peng:Yang:2021}.
In this paper, we focus on the frequentist MA approach with GAPLMs.

To obtain asymptotic optimality properties for MA estimators, the choice of weight is crucial. A common class of weight choice criteria in MA is based on cross-validation (CV). \cite{Hansen:Racine:2012} presented a jackknife MA estimator with the weights calculated by minimizing a leave-one-out (LOO) CV criterion for linear regression models. The same strategy was used for MA in high-dimensional linear regression \citep{Ando:Li:2014}, high-dimensional GLMs \citep{Ando:Li:2017}, longitudinal data models \citep{Gao:Zhang:Wang:Zou:2016} and vector autoregressive models \citep{Liao:Zong:Zhang:Zou:2019}.
There are other classes of criteria in the literature for the weight choice, see \cite{Hansen:2007}, \cite{Zhang:Yu:Zou:Liang:2016},  \cite{Zhang:Wang:2019}, \cite{Zhu:Wan:Zhang:2019}, and the references therein.

However, the MA methods with the LOO CV have heavy computational costs especially when the sample size is large. Thus, some attempts have been made to choose weights using the multifold CV criterion. For instance, \cite{Zhao2019Averaging} studied this alternative for multinomial and ordered logit models, and \cite{Liu:Long:Zhang:Li:2018} did it in a mixture copula framework. In this paper, to greatly reduce the computational cost, we also employ multifold CV criterion in the GAPLM framework.

To the best of our knowledge, few studies have been conducted on MA for GAPLMs.
An exception is \cite{Zhang:Liang:2011},  who presented a frequentist MA estimator for GAPLMs and provided some asymptotic results for a focused parametric function in a local misspecification framework. However, it is unknown whether the combined estimator has any asymptotic optimality property in terms of the statistical loss/risk. With the above background, in this paper, we shall study the proposed MA method with a data-driven weight choice for GAPLMs and establish an asymptotic optimality of the MA estimator.

In addition, if there are many candidate models, the computational burden will be heavy and the final MA estimator may perform poorly \citep{Zhang:Lu:Zou:2013}. In this case, model screening before using the MA may be helpful (see, e.g., \cite{Yuan:Yang:2005} and \cite{Zhang:Yu:Zou:Liang:2016}). Thus, in this paper, we will further provide an appropriate model screening procedure in conjunction with our proposed MA method for GAPLMs.

The main  contributions of our paper are as follows.
\begin{enumerate}
  \item We propose a computationally feasible multifold CV based MA procedure for GAPLMs. The asymptotic optimality of the proposed MA method is derived, that is, the multifold CV-based MA estimator achieves the lowest possible Kullback-Leibler (KL) loss asymptotically when all the candidate models
are misspecified.
  \item When the candidate model set includes the correct models, the over-consistency of the weights chosen by the multifold CV is proved, that is, the proposed multifold CV asymptotically assigns all weights to the correct models. As a by-product, based on the MA weights, we develop a variable importance measure to quantify the importance of the predictors in GAPLMs, which can asymptotically screen out all the variables in the true model.
  \item We develop a model screening approach based on the distance correlation, which is model-free. The advantages of this approach are further investigated in theory.
\end{enumerate}

The rest of this article is organized as follows. Section 2 provides a detailed introduction to our MA procedure for GAPLMs and also presents variable importance method, and the corresponding theoretical properties are presented in Section 3. In Section 4, we study the optimal MA procedure with an appropriate model screening step. Simulation studies on the proposed MA method are presented in Section 5. In Section 6, we study a real data example. Concluding remarks are offered in Section 7.  Additional numerical results and detailed proofs of the theorems are provided in the Supplementary Materials.

\section{Model Averaging and Variable Importance}
\label{sec:model}
Let $\{y_i\}_{i=1}^{n}$ be the independent and identically distributed random variables. We consider the exponential family:
$$
f(y_i|\theta_i,\phi)=\exp\left(\frac{y_i\theta_i-b(\theta_i)}{\phi}+c(y_i,\phi)\right), \quad i=1,\ldots,n.
$$
Here, $\theta_i$ is called the natural parameter, $\phi$ is the dispersion parameter, and $b(\cdot)$ and $c(\cdot,\cdot)$ are known functions. Let $\{x_i\}_{i=1}^{n}$ be $n$ nonstochastic predictor vectors with $x_i=(x_{i1},\ldots,x_{ip})^{\mathrm{T}}$.
The generalized additive partial linear model (GAPLM) can be written as
\begin{equation}\label{GAPLMs}
\eta_i\triangleq g(\mu_i)=\beta_0+\sum_{j=1}^{d_1}h_j(x_{ij})+\sum_{j=d_1+1}^{p}{\beta}_jx_{ij}, \quad i=1,\ldots,n,
\end{equation}
where $\mu_i=E(y_i)$, $g$ is a specified link function, $\beta_0$ is an intercept term, $h_j$ $(j=1,\ldots,d_1)$ is the unknown function, and ${\beta}_j$ $(j=d_1+1,\ldots,p)$ is the regression coefficient in the parametric part. For $j=1,\ldots,d_1$, suppose that each $x_{ij}$ takes values in $[a_j,b_j]$, where $a_j< b_j$ are finite numbers. The number of covariates in the parametric part is $d_2=p-d_1$.
Throughout this paper, we suppose that $h_j$'s are identifiable. The parameter $\theta_i$ is connected to $x_i$  by the form $\theta_i=\eta_i=\beta_0+\sum_{j=1}^{d_1}h_j(x_{ij})+\sum_{j=d_1+1}^{p}{\beta}_jx_{ij}$.

In this paper, we apply the polynomial spline based procedure to approximate the nonparametric functions in the GAPLM.
Let $\mathcal{S}_j$ be the space of the polynomial splines of degree $l_j\geq1$, and let $J_j$ be the number of interior knots in the interval $[a_j,b_j]$ for $j=1,\ldots,d_1$.  We assume that there is a normalized B-spline basis $\Phi_j(x)=(\phi_{j,1}(x),\ldots,\phi_{j,v_j}(x))^{\mathrm{T}}$ for
the spline space $\mathcal{S}_j$, where $v_j=l_j+J_j$, and that it is allowed to increase as $n$ increases. Under proper smoothness assumptions, $h_{j}$ can be well approximated by the following functions in $\mathcal{S}_j$:
$$
h_j(x)\approx \sum_{k=1}^{v_j}{\beta}_{j,k}\phi_{j,k}(x)={\beta}_{j}^{\mathrm{T}}\Phi_j(x) \quad j=1,\ldots,d_1,
$$
where ${\beta}_{j}=({\beta}_{j,1},\ldots,{\beta}_{j,v_j})^{\mathrm{T}}$ and ${\beta}_{j,k}$ is the spline coefficient. Hence, $\eta_i$ can be approximated by
$$
{\eta}_i \approx \beta_{0}+\sum_{j=1}^{d_1}{\beta}_{j}^{\mathrm{T}}\Phi_j(x_{ij})+\sum_{j=d_1+1}^{p}{\beta}_jx_{ij}=z_i^{\mathrm{T}}\beta, \quad i=1,\ldots,n.
$$
Here, $z_i=(1,\Phi_1^{\mathrm{T}}(x_{i1}),\ldots,\Phi_{d_1}^{\mathrm{T}}(x_{id_1}),x_{i(d_1+1)},\ldots,x_{ip})^{\mathrm{T}}$ and $\beta=(\beta_0,{\beta}_1^{\mathrm{T}},\ldots,{\beta}_{d_1}^{\mathrm{T}},{\beta}_{d_1+1},\ldots,$ ${\beta}_{p})^{\mathrm{T}}$.


\subsection{Candidate Models and Weight Choice}
\label{subsec:weight}

Suppose that we have $M$ candidate models. Let $M_{k1}$ and $M_{k2}$ be the sets that contain the indices of the predictors in the nonparametric and parametric parts, respectively, under the $k$th candidate model. Note that $M_k\triangleq M_{k1}\bigcup M_{k2}\subseteq \{1,\ldots,p\}$, and we assume in this paper that each candidate model contains the intercept term. Moreover, the dimension of the B-spline basis can be different for different candidate models. For ease of presentation, denote by  $\Phi_{j,k}(x)=(\phi_{j,1}^{(k)}(x),\ldots,\phi_{j,v_{j,k}}^{(k)}(x))^{\mathrm{T}}$ $(j\in M_{k1})$ the B-spline basis used to approximate $h_j(x)$ under the $k$th candidate model, and we allow $v_{j,k}$ to diverge. Furthermore, let $J_{j,k}$ be the number of interior knots for $j\in M_{k1}$. Let $l_{j,k}$ be the degree of B-spline for $j\in M_{k1}$ under the $k$th candidate model. Clearly, $v_{j,k}=l_{j,k}+J_{j,k}$. Note that we allow different candidate models to use different locations of interior knots. In short, the setting of our candidate models considers both the uncertainty of the covariates and that of the smoothness of the nonparametric functions in the GAPLMs.

Let $d_{1k}$ and $d_{2k}$ be the number of covariates in the nonparametric and parametric parts, respectively, under the $k$th candidate model. Let $p_k=d_{1k}+d_{2k}$ and $D_k=\sum_{j\in M_{k1}}v_{j,k}+d_{2k}+1$.
Based on the approximation of the B-spline basis, we can write:
$$
f(y_i|z_{ki},\beta_{M_k})=\exp\left(\frac{y_iz_{ki}^{\mathrm{T}}\beta_{M_k}-b(z_{ki}^{\mathrm{T}}\beta_{M_k})}{\phi}+c(y_i,\phi)\right),
$$
where $z_{ki}$ is a $D_k$-dimensional vector that includes $\Phi_{j,k}^{\mathrm{T}}(x_{ij})$ $(j\in M_{k1})$ and $x_{ij}$ $(j\in M_{k2})$, and $\beta_{M_k}$ is the regression coefficient vector of the $k$th candidate model. 
We assume in this paper that the dispersion parameter $\phi$ is known. The maximum likelihood estimator (MLE) of the regression coefficient is given by
\begin{equation}\label{betaMk}
  \hat{\beta}_{M_k}=\argmax_{\beta_{M_k}} \sum_{i=1}^{n}\log{f(y_i|z_{ki},\beta_{M_k})}.
\end{equation}
Furthermore, we denote $\hat{\eta}_{ki}=z_{ki}^{\mathrm{T}}\hat{\beta}_{M_k}$ for $i=1,\ldots,n$.

Next, we consider the MA estimator based on $M$ candidate models. Let $w_k$ be the weight of the $k$th candidate model, and let the weight vector $w=(w_1,\ldots,w_M)^{\mathrm{T}}$ belong to the set
\begin{equation}\label{weight set}
  \mathcal{W}=\bigg\{w \in[0,1]^{M}: \sum_{k=1}^{M}w_k=1\bigg\}.
\end{equation}
Then, the MA estimator of $\eta_i$ can be written as $\hat{\eta}_{i}(w)=\sum_{k=1}^{M}w_{k}\hat{\eta}_{ki}=\sum_{k=1}^{M}w_{k}z_{ki}^{\mathrm{T}}\hat{\beta}_{M_k}$. Furthermore, we formulate the MA estimator of $\mu_i$ as $\hat{\mu}_i(w)=b^{\prime}(\hat{\eta}_{i}(w))$ for $i=1,\ldots,n$.

{\rema{\label{condition}} \cite{Li2015flexible} and \cite{chen2018semiparametric} removed the restriction that the weights are non-negative and considered the weights $w_k\in \mathbb{R}$ $(k=1,\ldots,M)$ under the semiparametric model framework. However, they required that each candidate model includes only one covariate and that the covariates in different candidate models are mutually exclusive. In the context of linear regression, \cite{Hansen:Racine:2012} showed that the restriction $0\leq w_k\leq 1$ is a necessary condition for admissibility in the case of nested candidate models. Thus, we retain this restriction in this paper because the linear regression model is a special case of GAPLM and the candidate models are allowed to be nested.
}

A proper choice of weight is crucial in MA. We consider the CV method. As mentioned in \cite{Zhang:1993}, the performance of LOO CV in terms of the computational cost is unsatisfactory. This is especially the case for GAPLMs, since their estimation involves iterative optimalization, which brings about a heavy computational burden especially when the sample size is large. Thus, we use $K$-fold ($K>1$) CV to choose weights. Specifically,
the group $\{1,\ldots,n\}$ is divided into $K$ mutually exclusive subgroups $\mathcal{I}_1,\ldots,\mathcal{I}_K$. Let $|\cdot|$ be the number of elements of a set. For simplicity, we assume that each subgroup has the same number of elements, that is, $|\mathcal{I}_1|=\ldots=|\mathcal{I}_K|=m$ ($m\geq1$). 
Moreover, suppose that the division is as follows:
$$
\overbrace{1, \ldots, m}^{\mathcal{I}_{1}},  \overbrace{m+1, \ldots, 2 m}^{\mathcal{I}_{2}}, \ldots, \overbrace{Km-m+1, \ldots, K m}^{\mathcal{I}_{K}}.
$$
For $i=1,\ldots,K$, denote by $\mathcal{I}_i^{c}=\{1,\ldots,n\}\backslash \mathcal{I}_i$ the set whose elements are in the set $\{1,\ldots,n\}$ but not in the set $\mathcal{I}_i$. For the $k$th candidate model, let $\hat{\beta}_{k,\mathcal{I}_i^c}$ be the MLE of ${\beta}_{M_k}$ with the data corresponding to the subgroup $\mathcal{I}_i^c$, that is, $\{y_i,x_i\}_{i\in \mathcal{I}_i^c}$.

We propose the following $K$-fold CV based  weight choice criterion:
\begin{equation}\label{q-fold CV}
  C V(w)=\sum_{i=1}^{K}\sum_{j\in \mathcal{I}_i}\bigg(\phi^{-1}y_{j}\times\bigg(\sum_{k=1}^{M} w_{k} z_{k j}^{\mathrm{T}} \hat{\beta}_{k,\mathcal{I}_i^c} \bigg)-\phi^{-1}b\bigg(\sum_{k=1}^{M} w_{k} z_{k j}^{\mathrm{T}} \hat{\beta}_{k,\mathcal{I}_i^c}\bigg)+c(y_{j},\phi)\bigg).
\end{equation}
The resultant weight estimator  is
\begin{equation}\label{optimal w}
\hat{w}=\argmax_{w\in \mathcal{W}} CV(w),
\end{equation}
 where $\hat{w}=\left(\hat{w}_{1}, \ldots, \hat{w}_{M}\right)^{\mathrm{T}}$. When $m=1$, the criterion $CV(w)$ is reduced to the  LOO CV.
Furthermore, the resultant MA estimator of $\eta_i$ is given by $\hat{\eta}_i\triangleq\hat{\eta}_i(\hat{w})=\sum_{k=1}^{M} \hat{w}_{k} z_{k i}^{\mathrm{T}} \hat{\beta}_{M_k}$ for $i=1,\ldots,n$.

\subsection{Variable Importance Based on Model Averaging}

Variable importance is helpful to find the most important variables for understanding, interpretation, estimation, or prediction purposes. For instance, based on variable importance, decision-makers can change or replace variables, which may lead to cost and time savings in real data collection and analysis. Classically, one determines which variables are important based on
t-test values, regression coefficients, or p-values after obtaining a final linear regression model. However, these conventional methods not only cannot be applied to GAPLMs, but they also ignore variable selection uncertainty. In this subsection, we develop a reliable method of judging whether each of the regressors is important in GAPLMs based on our MA approach, called variable importance based on model averaging (VIMA). According to the candidate model set $\mathcal{M}=\left\{M_{k}, k=1, \ldots, M\right\}$ and the estimated weight $\hat{w}=\left(\hat{w}_{1}, \ldots, \hat{w}_{M}\right)^{\mathrm{T}}$ in \eqref{optimal w}, similar in spirit to \cite{Ye:Yi:Yang:2018}, we define the VIMA importance as follows.

{\defi{\label{importance}}The VIMA importance of the $j$th variable $X_{j}$, $j \in\{1, \ldots, p\}$, is defined as
$$
V_{j} \defby V(j ; \hat{w}, \mathcal{M})=\sum_{k=1}^{M} \hat{w}_{k} I\left(j \in M_{k}\right),
$$
where $I(\cdot)$ is an indicator function.
}

 The importance of the $j$th variable $X_{j}$ is the sum of the weights of the candidate models including the variable $X_{j}$, and $0\leq V_j \leq 1$. The theoretical property of VIMA depends on the behavior of the weights $\hat{w}$ in \eqref{optimal w}, which is explored next.

\section{Theoretical Results for Model Averaging and Variable Importance}
\label{sec:theory}
In this section, we establish the asymptotic optimality of the proposed MA method in the sense of minimizing the KL loss, and the over-consistency of the weights $\hat{w}$ and VIMA.
Let $y^{*}=(y_1^{*},\ldots,y_n^{*})^{\mathrm{T}}$ be an independent copy of $y=(y_1,\ldots,y_n)^{\mathrm{T}}$.
To measure the divergence between the MA estimator $\hat{\eta}_{i}(w)$ and the true parameter $\eta_i$, we consider the KL divergence, which is defined as
\begin{flalign}
KL(w)&= \sum_{i=1}^{n}E_{y^{*}}(\log(f(y_i^*|\eta_i,\phi))-\log(f(y_i^*|\hat{\eta}_i(w),\phi)))\nonumber\\
&=\sum_{i=1}^{n}\phi^{-1}\bigg(b^{\prime}(\eta_i)\times\bigg(\eta_i-\sum_{k=1}^{M} w_{k} z_{k i}^{\mathrm{T}} \hat{\beta}_{M_k} \bigg)-\bigg(b(\eta_i)-b\bigg(\sum_{k=1}^{M} w_{k} z_{k i}^{\mathrm{T}} \hat{\beta}_{M_k}\bigg)\bigg)\bigg),\label{KL loss}
\end{flalign}
and correspondingly, the expected KL divergence is defined as
$$
R(w)=E_y(KL(w)),
$$
where the expectation is calculated with respect to the joint density of $y=(y_1,\ldots,y_n)^{\mathrm{T}}$.

Let $D_{max}=\max_{1\leq k\leq M}D_k$ and $\eta=(\eta_1,\ldots,\eta_n)^{\mathrm{T}}$. Denote by $\beta_{M_k}^{0}$ the $D_k$-dimensional pseudo-true regression parameter vector that minimizes the KL distance measure between the true model and the $k$th candidate model. We assume that the pseudo-true regression parameter exists for each candidate model (see \cite{Ando:Li:2017} for theoretical support).
Let $\eta_{k}^{0}=(\eta_{k1}^{0},\ldots,\eta_{kn}^{0})^\mathrm{T}=(z_{k1}^{\mathrm{T}}\beta_{M_k}^{0},\ldots,z_{kn}^{\mathrm{T}}\beta_{M_k}^{0})^\mathrm{T}$ and $Z_{k}=(z_{k1},\ldots,z_{kn})^\mathrm{T}$.
For any non-empty set $\mathcal{I}\subset \{1,\ldots,n\}$, let $Z_{k,\mathcal{I}}$ be a $|\mathcal{I}|\times D_k$ submatrix of $Z_{k}$ that is obtained by extracting its
rows corresponding to the indices in $\mathcal{I}$. Let $\hat{\beta}_{k,\mathcal{I}}$ be the MLE of $\beta_{M_k}$ using the sample observations whose indices are in $\mathcal{I}$.
Moreover, let $\tilde{\eta}_{k}=(\tilde{\eta}_{k1},\ldots,\tilde{\eta}_{kn})^\mathrm{T}=((Z_{k,\mathcal{I}_1}\hat{\beta}_{k,\mathcal{I}_1^c})^{\mathrm{T}},\ldots,(Z_{k,\mathcal{I}_K}\hat{\beta}_{k,\mathcal{I}_K^c})^{\mathrm{T}})^\mathrm{T}$
and $\hat{\eta}_{k}=(\hat{\eta}_{k1},\ldots,$ $\hat{\eta}_{kn})^\mathrm{T}=((Z_{k,\mathcal{I}_1}\hat{\beta}_{M_k})^{\mathrm{T}}$ $,\ldots,(Z_{k,\mathcal{I}_K}\hat{\beta}_{M_k})^{\mathrm{T}})^\mathrm{T}$.
Furthermore, let $\bar{B}_k={\rm{diag}}\{b^{''}(z_{ki}^\mathrm{T}\bar{\beta}_{M_k})\}_{i=1,\ldots,n}$ be the $n\times n$ diagonal matrix for every $\bar{\beta}_{M_k}$ lying between $\hat{\beta}_{M_k}$ and ${\beta}^{0}_{M_k}$,
and let $\bar{B}_{k,\mathcal{I}}={\rm{diag}}\{b^{''}(z_{ki}^\mathrm{T}\bar{\beta}_{k,\mathcal{I}})\}_{i\in\mathcal{I}}$ be a $|\mathcal{I}|\times |\mathcal{I}|$ diagonal matrix for every $\bar{\beta}_{k,\mathcal{I}}$ lying between $\hat{\beta}_{k,\mathcal{I}}$ and $\hat{\beta}_{M_k}$.

\subsection{Asymptotic Optimality}

In this subsection, we consider the scenario in which all the candidate models are misspecified. In the following, unless otherwise stated, asymptotics are with respect to $n\rightarrow \infty$. Let $C$ be a generic positive constant.
To investigate the asymptotic behavior of the proposed MA method, we need the following conditions.
{\cond{\label{derivate}} $n^{-1}\sum_{i=1}^{n}b^{\prime}(\eta_i)^2\leq C< \infty$, $\sup_{w\in \mathcal{W}} n^{-1}\sum_{i=1}^{n}b^{\prime}(\bar{\eta}_i)^2=O_p(1)$, and $\sup_{w\in \mathcal{W}}$ $ n^{-1}\sum_{i=1}^{n}E(b^{\prime}(\bar{\eta}_i)^2)\leq C <\infty$ for every $\bar{\eta}_i$ lying between $\sum_{k=1}^{M} w_{k} \hat{\eta}_{ki}$ and $\sum_{k=1}^{M} w_{k} {\eta}_{ki}^0$. Moreover, $\sup_{w\in \mathcal{W}} n^{-1}\sum_{i=1}^{n}b^{\prime}(\tilde{\bar{\eta}}_i)^2=O_p(1)$ for every $\tilde{\bar{\eta}}_i$ lying between $\sum_{k=1}^{M} w_{k} \hat{\eta}_{ki}$ and $\sum_{k=1}^{M} w_{k} \tilde{\eta}_{ki}$.
}
{\cond{\label{Riesz}} Uniformly for the non-empty set $\mathcal{I}\in\{\mathcal{I}_1^c,\ldots,\mathcal{I}_K^c\}$,
$$
\begin{aligned}
&\lambda_{\max }\bigg(\frac{1}{|\mathcal{I}|} Z_{k,\mathcal{I}}^\mathrm{T} Z_{k,\mathcal{I}}\bigg)\leq C\times D_k \mbox{ and }
\max_{1\leq k\leq M} \lambda_{\max}\bigg(\bigg(\frac{1}{|\mathcal{I}|} Z_{k,\mathcal{I}}^\mathrm{T}\bar{B}_{k,\mathcal{I}} Z_{k,\mathcal{I}}\bigg)^{-1}\bigg)=O_p(1),
\end{aligned}
$$
where $\lambda_{\max }(\cdot)$ denotes the largest eigenvalue. Moreover,
$$
\begin{aligned}
&\lambda_{\max }\bigg(\frac{1}{n} Z_{k}^\mathrm{T} Z_{k}\bigg)\leq C\times D_k
\mbox{ and } \max_{1\leq k\leq M} \lambda_{\max}\bigg(\bigg(\frac{1}{n} Z_{k}^\mathrm{T}\bar{B}_{k} Z_{k}\bigg)^{-1}\bigg)=O_p(1).
\end{aligned}
$$
}
{\cond{\label{moment1}} Uniformly for $k\in \{1,\ldots,M\}$, $E\|D_{max}^{-1/2}n^{1/2}(\hat{\beta}_{M_k}-{\beta}_{M_k}^0)\|^{2}_2\leq C<\infty$ and $\max_{1\leq i\leq K}$ $\|D_{max}^{-1/2}(n-m)^{1/2}(\hat{\beta}_{k,\mathcal{I}_i^c}-{\beta}_{M_k}^0)\|_2^{2}=O_p(1)$.
}

{\cond{\label{fold and candi}} As $n\rightarrow\infty$, $D_{max}/n^{1/2}\rightarrow 0$ and $m/n^{1/2}\leq C<\infty$.
}
{\cond{\label{pseudo}} Uniformly for $k\in \{1,\ldots,M\}$, $n^{-1}\|\eta-\eta_{k}^0\|_2^2\leq C<\infty$ and $n^{-1}\|\eta_{k}^0\|_2^2\leq C<\infty$.
}
{\cond{\label{moment}} For some fixed integer $1\leq C^{*}< \infty$, $E(\varepsilon_{i}^{4 C^{*}}) \leq C<\infty, \; i=1, \ldots, n$, where $\varepsilon_i=y_i-\mu_i=y_i-E(y_i)$.
}
{\cond{\label{(A3)}}  As $n\rightarrow\infty$, for the fixed integer $C^*$ in Condition \ref{moment}, $M^{1/(2C^*)}D_{max}n^{{1}/{2}}/\xi_n\rightarrow 0$, where $\xi_n=\inf_{w\in \mathcal{W}}R(w)$.
}


Condition \ref{derivate} introduces some requirements for the first derivative of function $b(\cdot)$, which is similar to Condition (A4) in \cite{Ando:Li:2017} and to Condition C.8 in \cite{Zhang:Yu:Zou:Liang:2016}.
Condition \ref{Riesz} is commonly used in literature \citep{Fan:Peng:2004, Li:Lv:Wan:Liao:2020}.
The preceding part of Condition \ref{moment1} can be viewed as a variant of Condition C.10 in \cite{Zhang:Yu:Zou:Liang:2016}, which implies that $D_{max}^{-1/2}n^{1/2}(\hat{\beta}_{M_k}-{\beta}_{M_k}^0)$ has the finite second order moment, and the last part is frequently used in literature \citep{Liu:Long:Zhang:Li:2018}.
Condition \ref{fold and candi} restricts the maximum dimension of the candidate models, the smoothness of the nonparametric
functions in the candidate models, and the size of each fold.
The first part of Condition \ref{pseudo} reflects the degree of approximation between $\eta$ and $\eta_{k}^0$, which is used in \cite{Ando:Li:2017}, and the second part of Condition \ref{pseudo} is similar to Condition C.2 in \cite{Zhang:Yu:Zou:Liang:2016}.
Condition \ref{moment} is a common condition; see \cite{Wan:Zhang:Zou:2010}, and \cite{Zhang:Zou:Carroll:2015} for the linear model, and \cite{Ando:Li:2017} for the GLMs.
Condition \ref{(A3)} is the extension of Condition C.6 in \cite{Zhang:Yu:Zou:Liang:2016} under the situation  with the diverging  number of candidate models. If we suppose that $M$ is fixed or that Condition \ref{moment} holds for a sufficiently large $C^*$, then  Condition \ref{(A3)} is analogous to Condition C.6 in \cite{Zhang:Yu:Zou:Liang:2016}.

{\theo{\label{theorem1}} If Conditions \ref{derivate}--\ref{(A3)} are satisfied, then we have
$$
\frac{KL(\hat{w})}{\inf_{w\in \mathcal{W}}KL(w)} \rightarrow 1
$$
in probability as $n\rightarrow \infty$.
}

Theorem \ref{theorem1} shows that our MA estimator asymptotically achieves the smallest possible KL loss $\inf_{w\in \mathcal{W}}KL(w)$ based on the multifold CV criterion.

{\rema{\label{condition}} Condition \ref{(A3)} implies an upper bound on the number $M$ of the candidate models. Specifically,
if we assume that $\xi_n=\inf_{w\in \mathcal{W}}R(w)=\Omega(n)$, where $\xi_n=\Omega(n)$ states that there are some constants $c_1,c_2>0$ such that $c_1n<\xi_n<c_2n$, then we have $M^{1/(2C^*)}=o(n^{1/2}D_{max}^{-1})$ from Condition \ref{(A3)}. With Condition \ref{fold and candi}, if we further assume that $D_{max}=\Omega(n^{1/2-\vartheta})$ for $0<\vartheta\leq1/2$, then
 the order of $M$ is $o(n^{2C^*\vartheta})$.
}

\subsection{Over-consistency of Weights and VIMA}

  In this subsection,  we assume that each $h_j$ has continuous $r$th derivative that satisfies the Lipschitz condition of order $\upsilon$, where $r$ is a positive integer and $\upsilon\in (0,1]$. Let $q=r+\upsilon> 2$. Furthermore, suppose that the degree of the polynomial splines $l_{j,k}$ and the number of covariates are fixed for the $k$th candidate model. Moreover, we use equally spaced knots. For $1/(2q)<\gamma<1/4$, let the number of spline knots $J_{j,k}$ be of order $n^{\gamma}$ for the $k$th candidate model. Denote $\alpha=1/2-\gamma/2$.

In the following, suppose that not all predictors in the parametric part have contributions in predicting the response. Let  $\mathcal{I}_{T}=\{1,\ldots,d_1\}\bigcup \{j: \beta_j\neq 0\}$ and $\mathcal{I}_{F}=\{j: \beta_j=0\}$. To better define quasi-correct models, for each nonparametric component, we assume that the B-spline approximation of each $h_j(x_{ij})$ cannot be a linear function. The $k$th candidate model is called a quasi-correct model if $\mathcal{I}_{T} \subseteq M_k$ and the potential effects (nonparametric or parametric) of covariates $x_{ij}$'s $(j\in\mathcal{I}_{T})$ on $y_i$ are correctly specified, where $M_k$ is the set that contains the indices of covariates in the $k$th candidate model. Particularly, the $k$th quasi-correct model is a quasi-true model if $\mathcal{I}_{T} = M_k$. For the performance of the candidate model set, we consider the \emph{weak
inclusion} property.
{\defi{\label{inclusion}} A candidate model set is said to have the weak inclusion property if there is at least one quasi-correct model in the candidate model set.
}

 The \emph{weak inclusion}  property is a minor requirement for the candidate model set. Note that in Definition \ref{inclusion}, the quasi-true model is not necessarily in the candidate model set.
 Furthermore, denote by $\mathcal{I}_{\text{cor}}\subseteq \{1,\ldots,M\}$ the index set of the quasi-correct models in the candidate model set.
 By Taylor's expansion, we can calculate $KL(w)$ as follows:
 \begin{flalign}
 KL(w)&=\sum_{i=1}^{n}\phi^{-1}\bigg(b^{\prime}(\eta_i)\times\bigg(\eta_i-\sum_{k=1}^{M} w_{k} z_{k i}^{\mathrm{T}} \hat{\beta}_{M_k} \bigg)-\bigg(b(\eta_i)-b\bigg(\sum_{k=1}^{M} w_{k} z_{k i}^{\mathrm{T}} \hat{\beta}_{M_k}\bigg)\bigg)\bigg)\nonumber\\
 &=\sum_{k \in \mathcal{I}_{\text{cor}}}\sum_{i=1}^{n}w_k(\phi^{-1}b^{\prime}(\eta_i)\times(\eta_i-z_{k i}^{\mathrm{T}} \hat{\beta}_{M_k} )-\phi^{-1}(b^{\prime}(\bar{\eta}_i(w))(\eta_i-z_{k i}^{\mathrm{T}} \hat{\beta}_{M_k})))\nonumber\\
 &+\sum_{k \notin \mathcal{I}_{\text{cor}}}\sum_{i=1}^{n}w_k(\phi^{-1}b^{\prime}(\eta_i)\times(\eta_i-z_{k i}^{\mathrm{T}} \hat{\beta}_{M_k} )-\phi^{-1}(b^{\prime}(\bar{\eta}_i(w))(\eta_i-z_{k i}^{\mathrm{T}} \hat{\beta}_{M_k})))\nonumber\\
 &\defby KL^*_{1}(w)+KL^*_{2}(w), \label{R1R2}
 \end{flalign}
where $\bar{\eta}_i(w)$ is a point lying between $\eta_i$ and $\sum_{k=1}^{M} w_{k}z_{k i}^{\mathrm{T}} \hat{\beta}_{M_k}$. Denote $w_{\text{cor}}=\sum_{k\in \mathcal{I}_{\text{cor}}} w_k$ and $\hat{w}_{\text{cor}}=\sum_{k\in \mathcal{I}_{\text{cor}}} \hat{w}_k$.
Let $\tilde{\xi}_n=\inf_{w\in \widetilde{\mathcal{W}}}(1-w_{\mathrm{cor}})^{-1}R^{*}_{2}(w)$, where $R^{*}_{2}(w)=E_y(KL_2^{*}(w))$ and $\widetilde{\mathcal{W}}=\{w \in[0,1]^{M}: \sum_{k=1}^{M}w_k=1 \mbox{ and }\sum_{k\in\mathcal{I}_{\mathrm{cor}}}w_k\neq 1\}$.

  In addition, let $z_{ki,(j)}=\Phi_{j,k}^{\mathrm{T}}(x_{ij})$ for $j\in M_{k1}$ and $k\in \mathcal{I}_{\text{cor}}$, where $\Phi_{j,k}^{\mathrm{T}}(\cdot)$ is a B-spline basis function (see also Section 2.1 for details). Let $\hat{\beta}_{k}^{(j)}$ be the estimated regression spline coefficients that correspond to the nonparametric function $h_j(x)$ for $j\in M_{k1}$ and $k\in \mathcal{I}_{\text{cor}}$. For $j\in M_{k2}$ and $k\in \mathcal{I}_{\text{cor}}$, let $\hat{\beta}_{k}^{(j)}$ be the estimated regression coefficient of $\beta_j$. Note that $\hat{\beta}_{k}^{(j)}$ can be calculated by \eqref{betaMk}.
 Some conditions required for the over-consistency of the weights $\hat{w}$ in \eqref{optimal w} and the variable importance are as follows.
{\cond{\label{deri}} $\sup_{w\in \mathcal{W}}\sum_{i=1}^{n}n^{-1}b^{\prime}(\bar{\eta}_i(w))\leq C<\infty$  for every $\bar{\eta}_i(w)$ lying between $\eta_i$ and $\sum_{k=1}^{M} w_{k}z_{k i}^{\mathrm{T}} \hat{\beta}_{M_k}$.
}

{\cond{\label{convergence}} Uniformly for $k\in \mathcal{I}_{\mathrm{cor}}$, $n^{-1}\sum_{i=1}^{n}(\sum_{j=1}^{d_1}h_j(x_{ij})-\sum_{j\in M_{k1}}z_{ki,(j)}^{\mathrm{T}}\hat{\beta}_{k}^{(j)})^2=O_p(n^{-2\alpha})$, where $\alpha=1/2-\gamma/2$, and $n^{1/2}(\hat{\beta}^{(j)}_{k}-\beta_j)=O_p(1)$ for $j\in M_{k2}$.
}

{\cond{\label{uniformly integrable}} $\tilde{\xi}_n^{-1}\sup_{w\in \mathcal{W}}KL_{1}^{*}(w)$ is uniformly integrable.
}
{\cond{\label{nocandi}} As $n\rightarrow\infty$, $n^{1-\alpha}/\tilde{\xi}_n\rightarrow 0$ and $M^{1/2C^*}n^{1/2}/\tilde{\xi}_n\rightarrow 0$ for the fixed integer $C^*$ in Condition \ref{moment}, where $\tilde{\xi}_n=\inf_{w\in \widetilde{\mathcal{W}}}(1-w_{\mathrm{cor}})^{-1}R^{*}_{2}(w)$.
}

Condition \ref{deri} is similar to Condition \ref{derivate}. For an explanation of rationality about Condition \ref{convergence} see Remark \ref{rema convergence} below. Condition \ref{uniformly integrable} is similar to the condition of Theorem 3 in \cite{Zhang:Zou:Liang:Carroll:2020}, and it is only imposed to derive the order of its expectation.  
Condition \ref{nocandi} restricts the growth rate of $\tilde{\xi}_n$ and the number of candidate models.

{\rema{\label{rema convergence}}  For each $k\in\mathcal{I}_{\mathrm{cor}}$, under some regularity assumptions, \cite{Wang:Liu:Liang:Carroll:2011} showed that $n^{-1}\sum_{i=1}^{n}(\sum_{j=1}^{d_1}h_j(x_{ij})-\sum_{j\in M_{k1}}z_{ki,(j)}^{\mathrm{T}}\hat{\beta}_{k}^{(j)})^2=O_p(n^{-2\alpha})$ and $n^{1/2}(\hat{\beta}^{(j)}_{k}-\beta_j)=O_p(1)$ for $j\in M_{k2}$ (see Theorems 1 and 2 in \cite{Wang:Liu:Liang:Carroll:2011}). If the number of quasi-correct models is fixed, then Condition \ref{convergence} can be naturally satisfied. Thus, it is reasonable to assume that Condition \ref{convergence} holds uniformly for $k\in \mathcal{I}_{\mathrm{cor}}$.
}

{\theo{\label{consistency of w}} Under Conditions \ref{derivate}--\ref{moment} and Conditions \ref{deri}--\ref{nocandi}, if the candidate model set has the weak inclusion property, then we have $\hat{w}_{\mathrm{cor}}\xrightarrow{P}1$ as $n\rightarrow\infty$.
}

Theorem \ref{consistency of w} shows that the multifold CV can assign all the weights to the correct models asymptotically when the candidate model set has the \emph{weak inclusion} property. Furthermore, based on Theorem \ref{consistency of w}, we present the asymptotic behavior of importance $V_j$ of the $j$th variable.

{\coll{\label{consistency of Vj}} Under the conditions of Theorem \ref{consistency of w}, if the candidate model set has the weak inclusion property, then we have
\begin{equation}\label{Vj}
\min_{j\in\mathcal{I}_T}V_j\xrightarrow{P}1, \mbox{ and } \max_{j\in \tilde{\mathcal{I}}_{F}}V_j\xrightarrow{P}0, \mbox{ as } n\rightarrow\infty,
\end{equation}
where $\tilde{\mathcal{I}}_{F}=\{1,\ldots,p\}\setminus \tilde{\mathcal{I}}_{T}$ and $\tilde{\mathcal{I}}_{T}=\{j: j\in M_k \mbox{ and } k\in\mathcal{I}_{\mathrm{cor}}\}$.
}

Note that the variables whose indices belong to the set $\tilde{\mathcal{I}}_{F}$ are not included in any of the correct models. Corollary \ref{consistency of Vj} shows the over-consistency of VIMA; that is, the VIMA importance of each variable in the true model converges to 1 in probability and to 0 for each variable outside the
correct models  based on the candidate model set with the \emph{weak inclusion} property.
{\rema{\label{VIMA screen}} After determining through a preliminary analysis which covariates have potentially nonlinear effects, the total number of potential candidate models is $2^p-1$. When $p$ is large, the proposed MA method has a heavy computational burden and is even computationally infeasible. Based on the over-consistency of VIMA, we can use VIMA to screen out some uninformative covariates, and we can further reduce the number of potential candidate models. However, a disadvantage of VIMA is still its computational burden, since we need to calculate the weights of all the candidate models. Thus, to reduce the number of candidate models, we present next an alternative model screening method.
}

\section{Model Averaging with Model Screening}
\label{sec:screening}

\subsection{Model Screening based on Distance Correlation}

If the  number of candidate models is large, it imposes a heavy computational burden on our MA estimator and may also affect the final performance of the MA estimator due to the high complexity of the candidate models. In this case, it is sensible to apply a screening process to reduce the number of candidate models before implementing our MA method.

For GAPLMs, since the link functions and nonparametric additive components are involved, previous screening methods based on parametric linear models frameworks (e.g., using the marginal Pearson correlation) and GLMs frameworks (e.g., using the marginal likelihood estimator) cannot be used. Thus, we consider the distance correlation (DC) that allows for arbitrary regression relationships between the response and the predictors (see, e.g., \cite{Li:Zhong:Zhu:2012}). Next, an example is given to illustrate the superiority of the DC to the Pearson correlation in screening nonparametric components.

We consider the following linear model
\begin{equation}\label{DC example}
  Y=\beta_0+h(X_1)+\beta_2 X_2+\varepsilon.
\end{equation}
In this model, each $X_i$ ($i=1,2$) is generated independently from a normal distribution with a mean of zero and a variance $\sigma^2>0$. The random error $\varepsilon$ is generated independently from the standard normal distribution. We assume that $h(X_1)=\beta_1X_1^2$ ($\beta_1\neq 0$). Through a simple calculation, we can verify the Pearson correlation: $\mathrm{corr}(X_1,Y)=0.$
 Thus, screening methods using the marginal Pearson correlation exclude the truly important predictor $X_1$. In contrast, the DC between $Y$ and $X_1$ is non-zero, that is, $\mathrm{dcorr}(X_1,Y)\neq 0$. We can thus identify the important predictor $X_1$ that corresponds to the nonparametric components based on the DC.

Next, we demonstrate theoretically that the DC can distinguish between the truly important and unimportant predictors. Let  $\mathcal{I}_{T}=\{1,\ldots,d_1\}\bigcup \{j: \beta_j\neq 0\}$ and $\mathcal{I}_{F}=\{j: \beta_j=0\}$ be the true and false index sets, respectively. Denote by $\mathrm{dcorr}(X_j,Y)$ the true DC between each predictor $X_j$ and the response $Y$ for $j=1,\ldots,p$. Let $\varrho_j=\mathrm{dcorr}^2(X_j,Y)$ and $\hat{\varrho}_j$ be the estimator of $\varrho_j$ based on the sample $\{y_i,x_i\}_{i=1}^{n}$. The specific estimation formula is given in \cite{szekely2007measuring} (see also S.5 in Part E of the Supplementary Materials for details). The predictors with larger $\hat{\varrho}_j$'s are regarded as more important.

{\theo{\label{ranking consistency}} Under Condition A1 in S.5 in Part E of the Supplementary Materials, if $\min_{j\in \mathcal{I}_{T}}\varrho_j\geq \max_{j\in \mathcal{I}_{F}}\varrho_j+Cn^{-\gamma_1}$ for some constants $C>0$ and $0\leq\gamma_1<1/2$, and if $\log{p}=o(n^{1-2(\gamma_1+\gamma_2)})$ for any $0<\gamma_2\leq 1/3-2\gamma_1/3$ or $\log{p}=o(n^{\gamma_2})$ for any $1/3-2\gamma_1/3<\gamma_2<1/2-\gamma_1$, then we have
$$
P\Big(\max_{j\in \mathcal{I}_F}\hat{\varrho}_j > \min_{j\in \mathcal{I}_T}\hat{\varrho}_j\Big)\rightarrow 0 \quad \mbox{as} \quad n\rightarrow\infty.
$$
}

 Theorem \ref{ranking consistency} shows that DC can separate the informative predictors from the non-informative ones with a high probability. Motivated by this property of DC, we provide  a model screening method based on DC  for
GAPLMs,    called the distance correlation model screening (DCMS),  which is described as follows.
\begin{enumerate}
\item[(1)] Sort the predictors in descending order based on the estimated DC between the predictors and the response without using the approximation of polynomial splines.
\item[(2)] Identify the covariates that have potentially nonlinear effects through a preliminary analysis, such as a visualization or a subjective expert judgment, and thus separate the covariates into two subsets, one considered for nonparametric effects and the other for linear effects.
\item[(3)] Choose $p$ nested candidate models, where the $k$th $(k\leq p)$ candidate model contains  the first $k$ predictors according to the order in step (1). Then the parameters of each candidate model can be estimated by \eqref{betaMk}.
\end{enumerate}
The numerical performance of DCMS is explored in Section 5.

\subsection{Model Averaging with DCMS}

Denote by $\mathcal{M}=\{1,\ldots,M\}$ the set that contains the indices of the $M$ candidate models obtained from a preliminary analysis. For instance, we can use all-subset models as the candidate models. Let $\mathcal{M}_{DC}$ be the subset of $\mathcal{M}$ that contains the indices of $p$ nested candidate models obtained based on DCMS.
Correspondingly, denote by
$$
\mathcal{W}_{DC}=\bigg\{w \in[0,1]^{M}: \sum_{k\in \mathcal{M}_{DC}}w_k=1 \mbox{ and } \sum_{k\not\in \mathcal{M}_{DC}}w_k=0\bigg\}
$$
the weight vector set for the reduced candidate model set. It can be seen that $\mathcal{W}_{DC}\subset \mathcal{W}$, where $\mathcal{W}$ has been defined in \eqref{weight set}.
The weight estimator $\hat{w}^{d}=(\hat{w}_1^d,\ldots,\hat{w}_M^d)^{\mathrm{T}}$ based on the reduced candidate model set
can be obtained as follows: $$\hat{w}^d=\argmax_{w\in \mathcal{W}_{DC}} CV(w),$$ where $\hat{w}^d_k=0$ for $k\not\in \mathcal{M}_{DC}$.
Similar to Theorem \ref{theorem1}, we next establish Theorem \ref{theorem1.1} that gives the asymptotic optimality under the reduced candidate model set. We define the asymptotically lossless (ALL) property before establishing Theorem \ref{theorem1.1}.
{\defi{\label{ALL screening}} {\rm{(ALL property)}} A weight set $\mathcal{W}_S$ $(\mathcal{W}_S\in \mathcal{W})$ obtained by a model screening process is called ALL if there exist a weight vector sequences $\{w^n\}$ $(w^n=(w^n_1,\ldots,w^n_M)\in \mathcal{W}_S)$ such that $(KL(w^n)-\inf_{w\in\mathcal{W}}KL(w))/\xi_n \xrightarrow{P}0$ as $n\rightarrow\infty$, where $\xi_n=\inf_{w\in \mathcal{W}}R(w)$.}

The ALL property elaborates that the weight set $\mathcal{W}_S$ needs to include a weight vector $w^n$ that is good enough in the sense that the distance between $KL(w^n)$ and $\inf_{w\in\mathcal{W}}KL(w)$ is small relative to $\xi_n$. A similar property is also used in \cite{Zhang:Yu:Zou:Liang:2016}. In S.7 in Part E of the Supplementary Materials, we discuss in detail the rationality of the definition of the ALL property.



{\theo{\label{theorem1.1}} If Conditions \ref{derivate}--\ref{(A3)} are satisfied and $\mathcal{W}_{DC}$ satisfies the ALL property, then we have
$$
\frac{KL(\hat{w}^d)}{\inf_{w\in \mathcal{W}}KL(w)} \rightarrow 1
$$
in probability as $n\rightarrow \infty$.
}

Theorem \ref{theorem1.1} shows that the MA estimator obtained over the reduced candidate model set based on DCMS still asymptotically achieves the smallest possible KL loss $\inf_{w\in \mathcal{W}}KL(w)$ under some mild conditions.

{\rema{\label{ALL remark}} Theorem \ref{theorem1.1} relies on a good quality of the weight set $\mathcal{W}_{DC}$ obtained based on DCMS. We use a specific example to demonstrate that under some mild conditions, the ALL property can be achieved when the candidate model set includes some good models (see S.7 in Part E of the Supplementary Materials). Also, Theorem \ref{ranking consistency} guarantees that $\mathcal{M}_{DC}$ includes some good candidate models with a high probability. In addition, the numerical results in the simulation section illustrate that the proposed MA method with DCMS indeed performs very well.
}

\section{Simulation Studies}

In this section, we consider different simulation settings to evaluate the performance of the proposed MA method for GAPLMs in terms of the KL loss and the computing time in Section 5.1, and to evaluate the performance of the over-consistency of the weights and the variable importance in Section 5.2. We apply the cubic B-splines to approximate the additive functions, and we use the same number of spline knots for different additive functions in the same candidate model. Furthermore, denote by $J_{(k)}$ $(k=1,\ldots,M)$ the number of spline knots under the $k$th candidate model. In addition, we place the internal knots at the empirical quantiles \citep{Meier:2009High}. 

\subsection{KL Loss and Computing Time}
In this subsection, we denote our MA method as CV-$m$, where $m$ $(m\geq1)$ is the sample size of each fold. In our simulations, we implement three versions of CV-$m$ with $m=1$, $m=5$, and $m=10$, respectively.  For comparison, we also consider model selection methods (AIC and BIC) and MA approaches using the information criterion (SAIC and SBIC, \cite{Buckland:Burnham:Augustin:1997}).
Specifically, for the $k$-th candidate model, the weight assigned by SAIC is
$$
w_{\mathrm{AIC}, k}=\exp(-\mathrm{AIC}^{(k)} / 2) / \sum_{k=1}^{M} \exp (-\mathrm{AIC}^{(k)} / 2),
$$
where $\mathrm{AIC}^{(k)}=-2\sum_{i=1}^{n}\log(f(y_i|\hat{\eta}_{ki}))+2D_k$. The weights assigned by SBIC are defined similarly.
For the assessment of each approach, we use the KL-type loss defined as follows:
\begin{equation}\label{KL-type}
  KL_{\mathrm{loss}}=2n^{-1}\sum_{i=1}^{n}\phi^{-1}\left(b^{\prime}(\eta_i)\left(\eta_i-\hat{\eta}_{\mathcal{A},i} \right)-\left(b(\eta_i)-b\left(\hat{\eta}_{\mathcal{A},i}\right)\right)\right),
\end{equation}
where $\hat{\eta}_{\mathcal{A},i}$ is the estimator of $\eta_i$ obtained by the approach $\mathcal{A}$ $(\mathcal{A}\in\{\mbox{CV-1}, \mbox{CV-5}, \mbox{CV-10}, $ $\mbox{AIC}, \mbox{BIC}, \mbox{SAIC}, \mbox{SBIC}\})$.


We first apply the following logistic regression model:
\begin{flalign*}
\mathrm{logit\{Pr}(y_i = 1)\}=&\sum_{j=1}^{2}h_j(x_{ij})+\sum_{j=3}^{5}{\beta}_jx_{ij}\nonumber\\
=&\sin(2\pi x_{i1})+5x_{i2}^4+3x_{i2}^2-2+2{x}_{i3}+1.5{x}_{i4}-0.9{x}_{i5}
\end{flalign*}
for $i=1,\ldots,n$, where $x_{i1}$ and $x_{i2}$ are independently uniformly distributed on $[0,1]$, and $({x}_{i3},{x}_{i4},{x}_{i5})$ are generated from the multivariate normal distribution $N_{3}(0,\Sigma)$ with $\Sigma=(\rho^{|i-j|})_{3\times 3}$ and $\rho\in\{0,0.5,0.75\}$.

Next, we consider the uncertainty of the choice of covariates in Example 1 and the uncertainty of the degree of smoothness in Example 2.
We first restrict our attention to the uncertainty in the covariate set, that is, different candidate models include different covariates, but the number of spline knots used for each model is identical.
The settings of specific parameters are provided in Example 1.

\noindent\textbf{Example 1}.
We consider the sample size $n\in\{100,200\}$, the correlation $\rho\in\{0, 0.5, 0.75\}$, and $J_{(1)}=\ldots=J_{(M)}=[n^{1/5}]$, where $[x]$ denotes the smallest integer not less than $x$. With the five predictors, the number of candidate models is $M=2^{5}-1$.

The averaged $KL_{\mathrm{loss}}$ values of various methods (and their standard errors) over 500 replications for Example 1 are summarized in Table \ref{low_dim}. It can be seen that the CV-$m$ methods always yield smaller $KL_{\mathrm{loss}}$ values than their competitors in all situations.
Although the differences in the $KL_{\mathrm{loss}}$ values among all the methods becomes smaller as the sample size increases, the CV-$m$ methods still outperform the others. Furthermore, the standard errors of the CV-$m$ methods are relatively small compared with those of AIC, BIC, SAIC, and SBIC, which means the CV-$m$ methods are more stable.
The performance levels of CV-1, CV-5, and CV-10 are not very different in any setting, especially in the case of $n=200$.

\begin{table}[htbp]
\vspace{-0.5cm}
\begin{center}
 \caption{\label{low_dim}The averaged $KL_{\mathrm{loss}}$ values obtained using several methods and the
associated standard errors (in parenthesis) for Examples 1--3 ($\times 10^{-1}$).}
\small{
\begin{tabular}{lcccccccccccccccccc}
\hline\hline
\textbf{Example 1}          &AIC     &BIC      &SAIC      &SBIC      &CV-1      &CV-5      &CV-10 \\\hline
 $(n,\rho)=(100,0)$         &2.120   &1.848    &1.985     &1.713     &1.153     &1.171     &1.210 \\
                            &(0.071) &(0.055)  &(0.072)   &(0.047)   &(0.026)   &(0.027)   &(0.028)\\
 $(n,\rho)=(100,0.5)$       &2.134   &1.979    &2.005     &1.767     &1.116     &1.149     &1.178 \\
                            &(0.066) &(0.057)  &(0.068)   &(0.050)   &(0.021)   &(0.030)   &(0.048)\\
 $(n,\rho)=(100,0.75)$      &2.282   &2.011    &2.077     &1.756     &1.132     &1.137     &1.168  \\
                            &(0.079) &(0.052)  &(0.074)   &(0.044)   &(0.029)   &(0.029)   &(0.032)\\
 $(n,\rho)=(200,0)$         &0.817   &0.961    &0.788     &0.896     &0.599     &0.604     &0.606  \\
                            &(0.019) &(0.016)  &(0.018)   &(0.016)   &(0.011)   &(0.011)   &(0.011)\\
 $(n,\rho)=(200,0.5)$       &0.835   &0.998    &0.805     &0.916     &0.607     &0.609     &0.611     \\
                            &(0.018) &(0.016)  &(0.018)   &(0.015)   &(0.010)   &(0.010)   &(0.010)\\
 $(n,\rho)=(200,0.75)$      &0.797   &1.072    &0.776     &0.939     &0.589     &0.591     &0.595     \\
                            &(0.015) &(0.014)  &(0.015)   &(0.013)   &(0.010)   &(0.010)   &(0.010)\\
\hline\hline
\textbf{Example 2}          &AIC     &BIC      &SAIC      &SBIC      &CV-1      &CV-5      &CV-10 \\\hline
 $(n,\rho)=(100,0)$         &2.245   &1.291    &2.133     &1.246     &1.227     &1.117     &1.097 \\
                            &(0.118) &(0.038)  &(0.102)   &(0.038)   &(0.041)   &(0.027)   &(0.026)\\
 $(n,\rho)=(100,0.5)$       &2.357   &1.225    &2.200     &1.176     &1.211     &1.124     &1.133 \\
                            &(0.114) &(0.038)  &(0.102)   &(0.033)   &(0.035)   &(0.031)   &(0.040)\\
 $(n,\rho)=(100,0.75)$      &2.355   &1.324    &2.239     &1.283     &1.246     &1.156     &1.111  \\
                            &(0.109) &(0.039)  &(0.096)   &(0.042)   &(0.035)   &(0.030)   &(0.027)\\
 $(n,\rho)=(200,0)$         &0.890   &0.692    &0.815     &0.644     &0.636     &0.636     &0.638  \\
                            &(0.032) &(0.030)  &(0.014)   &(0.014)   &(0.015)   &(0.014)   &(0.015)\\
 $(n,\rho)=(200,0.5)$       &0.857   &0.689    &0.804     &0.638     &0.647     &0.644     &0.637     \\
                            &(0.027) &(0.014)  &(0.025)   &(0.014)   &(0.016)   &(0.015)   &(0.015)\\
 $(n,\rho)=(200,0.75)$      &0.789   &0.660    &0.728     &0.609     &0.600     &0.600     &0.599    \\
                            &(0.024) &(0.012)  &(0.020)   &(0.024)   &(0.013)   &(0.013)   &(0.013)\\
\hline\hline
\textbf{Example 3}          &AIC     &BIC      &SAIC      &SBIC      &CV-1      &CV-5      &CV-10 \\\hline
 $(n,\rho)=(100,0)$         &2.821   &2.045    &2.621     &1.943     &1.624     &1.591     &1.611 \\
                            &(0.122) &(0.088)  &(0.119)   &(0.087)   &(0.066)   &(0.058)   &(0.066)\\
 $(n,\rho)=(100,0.5)$       &3.027   &2.071    &2.729     &1.994     &1.501     &1.487     &1.499 \\
                            &(0.124) &(0.068)  &(0.113)   &(0.152)   &(0.033)   &(0.033)   &(0.032)\\
 $(n,\rho)=(100,0.75)$      &2.874   &2.268    &2.565     &2.045     &1.494     &1.410     &1.453  \\
                            &(0.132) &(0.079)  &(0.115)   &(0.050)   &(0.046)   &(0.035)   &(0.036)\\
 $(n,\rho)=(200,0)$         &1.002   &1.069    &0.947     &0.992     &0.756     &0.756     &0.760  \\
                            &(0.022) &(0.019)  &(0.021)   &(0.018)   &(0.014)   &(0.014)   &(0.014)\\
 $(n,\rho)=(200,0.5)$       &1.076   &1.276    &1.026     &1.150     &0.756     &0.754     &0.759     \\
                            &(0.032) &(0.029)  &(0.022)   &(0.028)   &(0.013)   &(0.012)   &(0.013)\\
 $(n,\rho)=(200,0.75)$      &1.067   &1.055    &0.978     &0.975     &0.742     &0.741     &0.744     \\
                            &(0.021) &(0.015)  &(0.021)   &(0.014)   &(0.012)   &(0.012)   &(0.012)\\
\hdashline
                            \multicolumn{8}{c}{The results of various methods based on the candidate models
with all subsets} \\\hdashline
$(n,\rho)=(100,0.75)$         &2.421   &1.805    &2.080     &1.593     &1.213     &1.204     &1.221 \\
                            &(0.161) &(0.088)  &(0.141)   &(0.083)   &(0.054)   &(0.052)   &(0.051)\\
 $(n,\rho)=(200,0.75)$       &1.030   &1.131    &0.935     &1.006     &0.698     &0.700     &0.701 \\
                            &(0.047) &(0.033)  &(0.044)   &(0.031)   &(0.025)   &(0.024)   &0.024)\\
\hline\hline
\end{tabular}}
\end{center}
\end{table}

\noindent\textbf{Example 2}. In this example, we consider the effect of the various numbers of B-spline knots on the MA methods. All the candidate models have the same covariates (i.e., $M_1=\ldots=M_M$ and $p_1=\ldots=p_k=p$), and $J_{(k)}\in\{1,\ldots,[(2n)^{1/5}]+2\}$ for $k=1,\ldots,M$. As a result, the total number of candidate models is $M=[(2n)^{1/5}]+2$.  The other parameters are set as in Example 1.

The simulation results for Example 2 are shown in Table \ref{low_dim} based on 500 simulation runs. It is seen from Table \ref{low_dim} that CV-5 and CV-10 perform better than AIC, BIC, SAIC, and SBIC in the overwhelming majority of cases. Although the performance of SBIC is close to that of CV-1, CV-1 is the relatively better method in most cases. In addition, BIC can also obtain small $KL_{\mathrm{loss}}$ values when $n=200$, but its performance is still worse than that of the CV-$m$ methods.
It is worth mentioning that the performance of CV-1 is inferior to that of CV-5 and CV-10 in the case of $n=100$. Yet, as the sample size increases, no significant difference is seen among CV-1, CV-5, and CV-10.

Next, to implement the screening strategy that we proposed in Section 4, we use the logistic regression model:
\begin{flalign*}
\mathrm{logit\{Pr}(y_i = 1)\}=&\sum_{j=1}^{2}h_j(x_{ij})+\sum_{j=3}^{9}{\beta}_jx_{ij}\nonumber\\
=&\sin(2\pi x_{i1})+5x_{i2}^4+3x_{i2}^2-2+2{x}_{i3}+1.5{x}_{i4}-0.9{x}_{i5}+0.05x_{i7}+0.08x_{i8}
\end{flalign*}
for $i=1,\ldots,n$, where the coefficients of $x_{i6}$ and $x_{i9}$ are equal to zero, $x_{i1}$ and $x_{i2}$ are independently uniformly distributed on $[0,1]$, and $({x}_{i3},\ldots,{x}_{i9})$ are generated from the multivariate normal distribution $N_{7}(0,\Sigma)$ with $\Sigma=(\rho^{|i-j|})_{7\times 7}$ and $\rho\in\{0,0.5,0.75\}$.

\noindent\textbf{Example 3}. The sample size $n$, the correlation $\rho$, and the number of spline knots $J_{(k)}$ $(k=1,\ldots,M)$ are set as in Example 1. With nine predictors, there are $2^9-1=511$ candidate models, which impose a heavy computational burden on our MA estimator. Thus, we adopt the DCMS to screen models prior to MA.

The simulation results for Example 3 are shown in Table \ref{low_dim}. It can be seen that CV-1, CV-5, and CV-10 perform much better than AIC, BIC, SAIC, and SBIC in all the cases. Specifically, CV-1, CV-5, and CV-10 always yield smaller $KL_{\mathrm{loss}}$ values and the standard errors than their competitors. Also, CV-1, CV-5, and CV-10 perform with almost indistinguishable differences. We further present the simulation results of various methods based on the all-subset models in the cases of $(n,\rho)=(100,0.75)$ and $(n,\rho)=(200,0.75)$ in Table \ref{low_dim}. The results show that the superiority of CV-1, CV-5, and CV-10 over AIC, BIC, SAIC, and SBIC is still apparent in this scenario. We also find that the CV-$m$ based on the all-subset candidate models delivers smaller averaged $KL_{\mathrm{loss}}$ values than the CV-$m$ with DCMS for each $m\in\{1,5,10\}$, but the absolute difference between their $KL_{\mathrm{loss}}$ values gradually approaches zero with the increase of the sample size. In other words, there is no significant
difference between the proposed MA method with DCMS and that without model screening concerning the averaged $KL_{\mathrm{loss}}$ values when the sample size is large.

Apart from the KL loss, we also consider the performance of CV-1, CV-5, and CV-10 from the perspective of the computational time under the settings of Examples 1--3. Detailed results are provided in Part A of the Supplementary Materials.

\subsection{Over-consistency of Weights and Variable Importance}
We next consider several simulation settings to evaluate the performance of the proposed VIMA method and the over-consistency of the weights $\hat{w}$ in \eqref{optimal w} when the candidate model set has the \emph{weak inclusion} property. To save space, the details are included in Part B of the Supplementary Materials.

\section{Application to the Vehicle Silhouettes Data}

In this section, we apply our MA method to the vehicle silhouettes data from the R package \texttt{mlbench}, which has been studied by \cite{Hsu:Lin:2002} and \cite{Li:Lv:Wan:Liao:2020}.
The dataset consists of 429 observations of different vehicle types, including the Opel Manta 400 (opel) with 212 vehicles and the Saab 9000 (saab) with 217 vehicles. Thus, the response $Y$ indicates the type of vehicle. It is equal to 0 (or 1) if the vehicle type is opel (or saab).
The corresponding covariates are 18 vehicle silhouette features that are summarized in Table S3 in Part C of the Supplementary Materials, and these covariates are continuous.
Based on a visualization (see Figure S2 in Part D of the Supplementary Materials), we find that the effects of $X_1,\ldots,X_6$ on the log odds seem potentially nonlinear.
Thus, we consider the following GAPLM:
\begin{equation}\label{logit real}
\mathrm{logit\{Pr}(Y= 1)\}=\sum_{j=1}^{6}h_j(X_{j})+\sum_{j=7}^{18}{\beta}_jX_{j},
\end{equation}
where $X_1,\ldots,X_6$ are set as nonparametric components. Note that all the covariates are standardized with a mean of 0 and a variance of 1. In the following, we model the effects of $X_1,\ldots,X_6$ using cubic B-splines with equidistant knots and with the number of knots set at $[(2n)^{1/5}]-1$ \citep{Huang:Yang:2004,Liao:Wan:He:Zou:2021}.

To assess the performance of each MA method, we use the following KL-type loss function:
\begin{equation}\label{KL data}
 KL_{\mathrm{real}}=-2 n_{test}^{-1}\sum_{i=1}^{n_{test}}\log f(y_{test,i}|\hat{\eta}_{test,i}),
\end{equation}
 where $y_{test,i}$ is the $i$th testing observation, $\hat{\eta}_{test,i}=\sum_{k=1}^{M} \hat{w}_{k} z_{test,k i}^{\mathrm{T}} \hat{\beta}_{M_k}$, $z_{test,k i}$ is the $i$th testing observation for the $k$th candidate model, $n_{test}$ is the number of observations in the testing data, and $\hat{w}_{k}$ and $\hat{\beta}_{M_k}$ are calculated based on the training data. Next, we implement the MA procedures by establishing the logistic model \eqref{logit real} for the vehicle data.

If we consider all the subset models as the candidate models, then there are $2^{18}-1$ submodels. Thus, to reduce the computational burden, we use the DCMS approach introduced in Section 4 to screen the candidate models.
Furthermore, we first randomly use 150 observations from the vehicle data to fit the model and then use the rest of the data as the test set. By repeating the above steps 500 times, we can obtain the mean KL-type loss defined in \eqref{KL data}. The results are shown in Table \ref{low-high data}.

\begin{table}[htbp]
\begin{center}
 \caption{\label{low-high data} The mean KL-type loss and averaged computational time (in seconds) for the vehicle silhouettes data.}
\small{
\begin{tabular}{ccccccccccccccccccc}
\hline\hline
              &AIC      &BIC       &SAIC      &SBIC      &CV-1      &CV-5      &CV-10
\\\hline
Loss           &1.723    &1.415     &1.673     &1.399     &1.239     &1.240     &1.241    \\
                  &(0.022)  &(0.003)  &(0.022)   &(0.003)   &(0.004)   &(0.004)   &(0.004)\\\hline
Time              &--       &--        &--        &--        &8.367     &2.016     &1.230  \\
                  &--       &--        &--        &--        &(0.028)   &(0.005)   &(0.003)\\
\hline\hline
\end{tabular}}
\end{center}
\footnotesize{Note: The standard errors are given in parenthesis, and ``--" means that the method is not used for the calculation.}
\end{table}

It is observed from Table \ref{low-high data} that the CV-$m$ methods perform better than AIC, SAIC, BIC, and SBIC.
The performance results of CV-1, CV-5, and CV-10 are quite close. Moreover, we summarize the averaged computational time (in seconds) of CV-1, CV-5, and CV-10 in Table \ref{low-high data} after 500 replications. Note that the computational time of  AIC, SAIC, BIC, and SBIC are not shown due to their poor performance with respect to the KL-type loss.  In terms of the computational efficiency, the CV-1 method has an unsatisfactory performance. Thus, the CV-5 and CV-10 methods are more recommendable in practice. In summary, this data analysis supports the CV-$m$ approaches extremely well in terms of both the KL-type loss and the computational feasibility.

\section{Conclusion}

In the context of GAPLMs, we proposed an MA approach based on multifold CV for computational ease. The asymptotical optimality of our method is derived in the sense of minimizing the KL loss when all the candidate models are misspecified, and the over-consistency of the weights chosen via multifold CV holds when the candidate model set has the \emph{weak inclusion} property.  As a by-product, we developed the VIMA procedure to evaluate variable importance, and showed that the significant predictors can be extracted effectively with VIMA. In addition, to address the severe computational burden caused by an excessive number of candidate models, we presented an appropriate model screening method and obtained the asymptotical optimality based on the reduced candidate model set under some conditions.
In comparison with AIC, BIC, SAIC, and SBIC, our numerical results show clear advantages of our proposed MA methods.

The current MA method developed in this paper works in the case $p<n$. The method and theory can be extended to the case $p>n$. In practice, the choice of the number of multifold $K$ is very important and a data-driven choice may be preferred, which is an interesting topic for future study.

\bigskip
\begin{center}
{\large\bf SUPPLEMENTARY MATERIAL}
\end{center}

\begin{description}

\item[Text document:] Supplemental Materials for ``Multifold Cross-Validation Model Averaging for Generalized Additive Partial Linear Model". (.pdf file)

\item[R-package for model averaging:] The package model averaging containing code to perform the diagnostic methods described in the article. (GNU zipped tar file)

\end{description}

\bibliographystyle{cheng}
\bibliography{ref}
\end{document}